# Inventions on GUI aesthetics
## A TRIZ based analysis

**Umakant Mishra**

Bangalore, India

http://umakantm.blogspot.in

**Contents**

## 1. Introduction

Aesthetics or "look and feel" is one of the most important features of any graphical user interface. Better aesthetics makes the interface user-friendlier and more popular. Better aesthetics helps the user to understand the meaning of various components and memorize the navigation paths. A better look and feel ultimately makes a GUI more efficient and effective. Various methods are adopted to improve the aesthetics of a GUI, such as;

- ⇛ By using colors
- ⇛ By using more dimensions
- ⇛ By using pictorial icons
- ⇛ By using (stereophonic) sound
- ⇛ By using better layout, and so on.

The developer faces various limitations while implementing aesthetics in a GUI. For example, the pictorial icons are more attractive and meaningful but they occupy more screen space. As the screen space is limited, it restricts the usage of large pictorial icons. One option may be to increase the screen size by implementing a large virtual desktop or multiple virtual desktops. But it is difficult for the user to navigate such virtual desktops.

A three-dimensional view may be more attractive and appropriate in some situations but implementing a three-dimensional view is difficult in a two dimensional workspace.

It is important to provide links to all the features on the desktop or a quick access panel. But too many icons or buttons sometimes creates confusion. For example, the early models of VCRs had a large number of buttons on their faceplate, which although attractive were causing confusion. Hence the later model of VCRs had only a few essential buttons on their faceplate and all other features were cleverly put somewhere behind a sliding door accessible only when required.

Thus it is important to restrict the temptation of putting everything on the first screen or load the rarely used buttons on the toolbar. One should ensure that the aesthetics of a GUI is not compromising with its accessibility and other important features.

## 2. Inventions on GUI aesthetics

### 2.1 Generating 3 Dimensional effects in 2 Dimensional Graphical User Interface (Patent 5864343)

**Background problem**

It is easy to achieve 3D effects in 3D graphics computer system by rendering objects on the 2D raster display using perspective algorithm. But 3D effects are not realistically achieved in 2D graphics computer systems. The problem is how to render 3D effects in a 2D raster display?

**Solution provided by the invention**

Naughton et al. invented a method (Patent 5864343, assigned by Sun Microsystems, Jan 99) to achieve 3D effects in 2D graphics computer systems. Normally the 2D graphics use only two coordinates (x and y values). The 3D effect can only be achieved by the graphics application provided each object is rendered with a relative depth (z value).

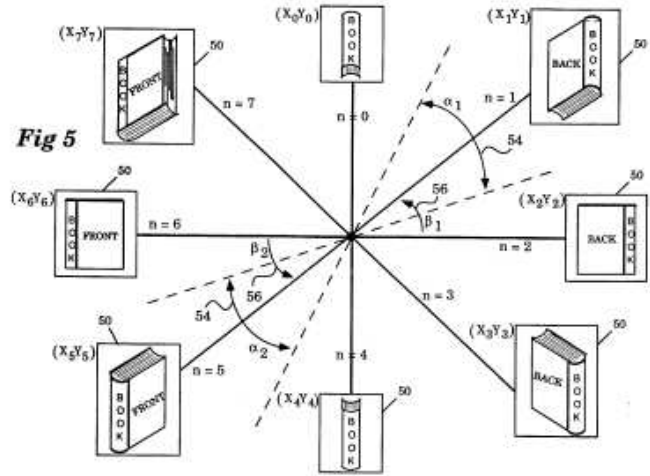

According to the invention, when the user moves the objects, the graphic toolkit routine re-computes the x, y and z coordinate values and displays objects to give a 3D movement. For example, the objects that are farther from the user will move slower than the objects that are closer to the user. The invention also uses stereophonic sound to enhance the 3D appearance of the objects displayed.

**TRIZ based analysis**

The invention adds a depth value (z) to the two dimensional display (x, y) of the objects which thereby produces 3D effect (Principle-17: Another dimension).

The method provides a stereophonic sound to enhance the 3D appearance of the objects displayed (Principle-38: Enrich).

**2.2 Improved GUI with Anthropomorphic characters (Patent 5886697)**

**Background problem**

Most popular graphical user interfaces are based on a "desktop metaphor" system, where the display screen is treated as the desktop and graphical symbols on the desktop represent environmental objects. But they cannot display the depth value of the tree structures or real world objects. There is a need for a better display system to control the real world devices.

**Solution presented by the invention**

Naughton et al. invented an intuitive graphical user interface (Patent 5886697, assigned by Sun Microsystems, Mar 99) to control remote devices from a hand held device. The GUI is organized in a geographic map structure that is simple to navigate as each space is represented by a familiar background image and familiar geographic surroundings. For example the background of a living room would contain the items typically found in a living room such as a lamp, a chair and a table.

According to the invention, the graphic objects on the screen represent graphic devices on the real world. To control the remote device a user should select the graphic object and control the device through the interface. The device is controlled by a device driver, which is located either within the remote device or within the handheld device depending on the remote device' sophistication.

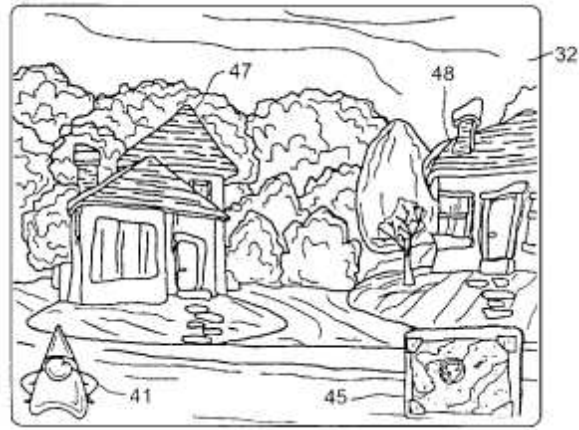

**TRIZ based analysis**

The invention uses colored cartoon-like icons representing actual objects and graphic image of geographical space for identifying the location of the objects more easily (Principle-26: Copying).

**2.3 Aspect and style elements of an improved graphical user interface (Patent 6005566)**

**Background problem**

There are various types of objects in a graphical user interface, such as, menus, toolbars, buttons, files, folders and so on. The prior art navigation tools can open an object and display various types of information about that object. Some objects may have different types of information, which are displayed in separate windows, such as, the classes of an object in one window, the methods of a selected class in another window and the source code for a selected method in another window. But all these methods display a predetermined type of information about an object. It is necessary to flexibly display various types of information associated with an object depending on the varied meaning of "opening" as the user desires.

**Solution provided by the invention**

Patent 6005566 (invented by Jones et al., assigned by Apple Computer Inc., issued in Dec 1999) disclosed an interactive graphical user interface having the capability to flexibly display various types of information associated with an object. The aspect element controls the particular type of information displayed for an object on a portion of a window, called a pane, while the style element controls the types of information displayed for all objects within that pane.

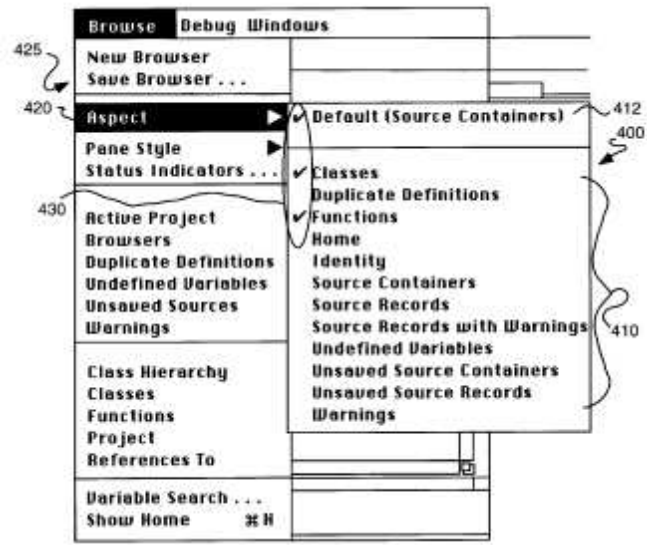

The invention displays an aspect menu, which contains elements for aspect settings. The user can manipulate the aspect settings to view any type of information that is linked to an object. For example, upon selecting a particular object, such as a vehicle, the user may wish to see a picture of a particular model of that vehicle, the options available for that model and descriptions of a particular available option.

**TRIZ based analysis**

The aspects of an object and styles of panes allow the user to customize viewing any type of information available for that object (Principle-15: Dynamize).

**2.4 System and method for customizing appearance and behavior of graphical user interfaces (6104391)**

**Background problem**

With the advantage of multitasking every user wants to run multiple applications each having their windows opened on the desktop. But the GUI of each different application has its own appearance and behavior. There may be dissimilarity in appearance and behavior between applications, which can be annoying and confusing to a user.

It would be desirable for the application developers and application users to have additional flexibility to provide greater control over the appearance and behavior of the desktop objects.

**Solution provided by the invention**

Patent 6104391 (invented by Jr. Johnston et al., Assigned by Apple Computer Inc, issued in Aug 2000) provides an appearance management layer that gives users (both developers and end users) the ability to customize the appearance and behavior of the desktop.

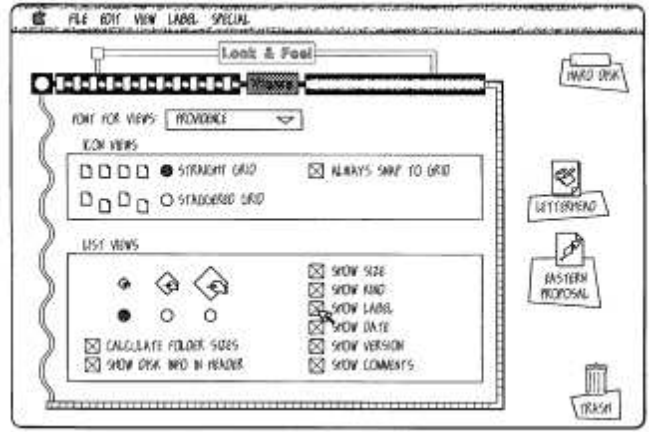

According to the invention the layer is provided between all the applications and the graphic subsystem, which actually writes to the display. In this way, a level of abstraction is provided between the client and the system so that the user switches between themes to customize the visual appearance and behavior even at runtime.

**TRIZ based analysis**

The user can customize the application and behavior of all the applications (Principle-15: Dynamize) in order to bring consistency of look and feel between different applications (Principle-6: Universality).

The invention provides a layer between the applications and the graphic subsystems. The customization is done at this layer (Principle-30: Thin and flexible).

**2.5 Advanced graphics controls (Patent 6177945)**

**Background problem**

A graphical user interface has various controls such as buttons, scrollbars, checkboxes, radio buttons, list boxes and so on. Although these existing graphic controls are functionally adequate for the current purposes, they will not be suitable for set-top boxes or Internet-enabled televisions as the user will expect more colorful and noticeable graphic controls. It is necessary to provide custom controls that can be supplied over relatively slow communication channels and utilized by unsophisticated computer devices.

**Solution provided by the invention**

Patent 6177945 (invented by Pleyer, assigned by Microsoft corporation, issued in Jan 2001) discloses a graphic control for an interactive user interface. The graphic components have both transparent portions and non-transparent portions. The components include a background component, a face component, a frame component and a focus component.

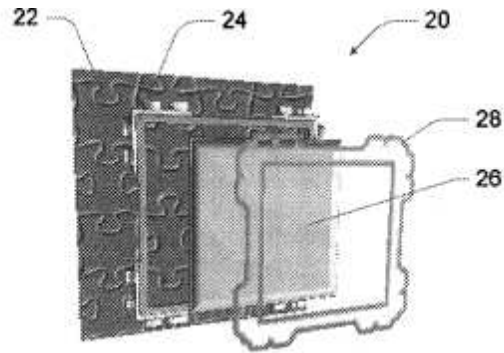

All components other than the background component are constructed of a texture bitmap and one or more luminance bitmaps. The texture bitmaps are tiled to conserve memory and data bandwidth. The luminance bitmaps can be supplied in a small size, and scaled using a special tiling technique.

**TRIZ based analysis**

The graphical components have transparent portions. They comprise of a texture element and a corresponding luminance element (Principle-32: Color change).

The graphic controls have layered structure of four components, viz., background component, face component, frame component and focus component, which are overlaid on each other (Principle-7: Nested doll).

Each graphic component consists of a texture bitmap and a luminance bitmap to give the desired effect (Principle-5: Merging).

**2.6 Pattern and color abstraction in a graphical user interface (Patent 6239795)**

**Background problem**

There are various elements of a graphic user interface such as windows, menus, scroll bars, toolbox etc. Each application developer can define his own nonstandard controls and window types as desired. Consequently there may be three applications running on a desktop each having windows of different look and feel. This dissimilarity in appearance and behavior between applications can be annoying and confusing to a user.

**Solution provided by the invention**

Patent 6239795 (invented by Ulrich et al., assigned by Apple Computer Inc., issued in May 2001) provides increased flexibility and control over the appearance and behavior of a user interface. According to the invention, sets of objects are grouped into themes to provide a user with a distinct overall impression of the interface. The user can switch between the themes, even at runtime, to change the desired appearance and behavior.

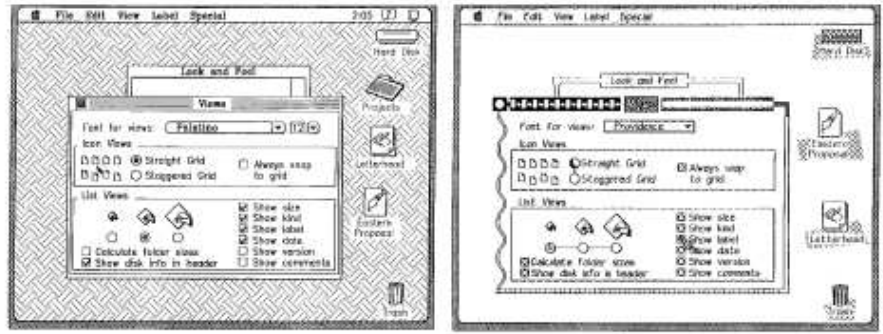

### TRIZ based analysis

According to the invention the appearance and behavior of objects are grouped into themes. (Principle-5: Merging).

The user can choose alternative themes to change the appearance and behavior of the graphic user interface (Principle-15: Dynamize).

### 2.7 Three-dimensional GUI windows with variable-speed perspective movement (Patent 6344863)

**Background problem**

With the improvement of graphic processing systems, the 3 dimensional pictures are becoming more commonplace. This makes the GUI interfaces to move from 2D to 3D perspective. The windows in a 3D environment are placed in front or behind one another instead of simply overlapping as in conventional 2D systems. Further the 3D desktop can be "rotated" on the computer display. The problem is how to implement a 3D interface on 2D screens.

**Solution provided by the invention**

Patent 6344863 (invented by Capelli et al., assigned by IBM, issued in Feb 2002) discloses a 3D graphical user interface wherein the windows have a "thickness" property.

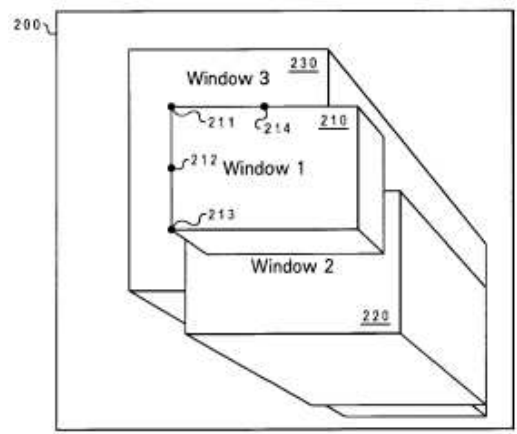

**TRIZ based analysis**

The invention adds a third dimension to the current 2D graphical user interface (Principle-17: Another dimension).

## 3. Summary

All developers feel that aesthetics is one of the important aspects of GUI, which needs to be addressed sincerely. Various inventions have tried to improve the aesthetics in different ways, such as,

- ⇛ By improving the look and feel of the graphical user interface
- ⇛ By bringing consistency in the look and feel of different GUI interfaces.
- ⇛ By providing features to adjust the look and feel by the user.
- ⇛ By providing features to change look and feel at runtime.

The inventions illustrated above are selected from US patent database. There are many more inventions too which try to improve the aesthetic aspect of Graphical User Interfaces.